\documentclass[preprint,12pt]{elsarticle}

\usepackage[utf8]{inputenc}
\usepackage{ragged2e}
\usepackage{bbm}
\usepackage{amsmath}
\usepackage{graphicx}
\usepackage{subcaption}
\usepackage{verbatim}
\usepackage{booktabs}
\usepackage{lineno}
\usepackage{amssymb}
\usepackage{float}
\usepackage{xcolor}
\usepackage[utf8]{inputenc}
\usepackage{amssymb}
\usepackage{appendix}
\usepackage[ruled,vlined]{algorithm2e}
\usepackage{algorithmicx}
\usepackage{setspace}
\usepackage{hyperref}

\newtheorem{remark}{Remark}%
\newcommand{\R}{\mathbb R}

\newcommand{\bsx}{\boldsymbol{x}}
\newcommand{\bsX}{\boldsymbol{X}}
\newcommand{\bsu}{\boldsymbol{u}}
\newcommand{\bsU}{\boldsymbol{U}}
\newcommand{\bsa}{\boldsymbol{a}}
\newcommand{\bsA}{\boldsymbol{A}}

\newcount\Comments  
\newcommand{\kibitz}[2]{\ifnum\Comments=1\textcolor{#1}{#2}\fi}

\begin{document}
\begin{frontmatter}

\title{ Enhanced sequential directional importance sampling for structural reliability analysis }

\author{Kai Cheng$^{a*}$,  Iason Papaioannou$^{a}$ and Daniel Straub$^{a}$ }

\affiliation{organization={Engineering Risk Analysis Group, Technical University of Munich},
 addressline={Theresienstr. 90}, 
city={ Munich},
 postcode={80333}, 
 country={Germany}}
 
\begin{abstract}

Sequential directional importance sampling (SDIS) \cite{cheng2023rare} is an efficient adaptive simulation method for estimating failure probabilities. It expresses the failure probability as the product of a group of integrals that are easy to estimate, wherein the first one is estimated with Monte Carlo simulation (MCS), and all the subsequent ones are estimated with directional importance sampling. In this work, we propose an enhanced SDIS method for structural reliability analysis. We discuss the efficiency of MCS for estimating the first integral in standard SDIS and propose using Subset Simulation as an alternative method. Additionally, we propose a Kriging-based active learning algorithm tailored to identify multiple roots in certain important directions within a specificed search interval. The performance of the enhanced SDIS  is demonstrated through various complex benchmark problems. The results show that the enhanced SDIS is a versatile reliability analysis method that can efficiently and robustly solve challenging reliability problems.  
 
\end{abstract}


\end{frontmatter}

\section{Introduction}

The fast evolution of computing power has enabled the development of various high-fidelity numerical models to simulate the performance of complex engineering systems in nearly all fields of engineering and science. Nevertheless, due to inevitable uncertainties involved in design, construction and operational conditions, the response of engineering systems remains uncertain. Reliability analysis is therefore an essential tool to assess the safety of engineering structures. It is concerned with estimating the failure probability 
 \begin{equation}
 P_f = \int_{g(\bsx)\leq 0} f_{\bsX}(\bsx)d\bsx = 
 \int_{\mathbb{R}^{n}} I\left(g(\bsx)\leq 0\right)f_{\bsX} (\bsx)  d\bsx = \mathbb{E}_{f_{\bsX}}\left[I\left(g(\bsX)\leq 0\right)\right],
 \label{eq:Pf_x}
 \end{equation}
where $\bsX =(X_1,...,X_n)^{\rm T}$ is the $n$-dimensional continuous input random vector with joint probability density function (PDF) $f_{\bsX} (\bsx)$, $g(\bsx)$ is the limit state function (LSF), and $I(\cdot)$ is the indicator function, which gives 1 if failure occurs, i.e., if $g(\bsx)\leq 0$, and 0 otherwise. $\mathbb{E}_{f_{\bsX}}[\cdot]$ denotes the expectation operator with respect to the density $f_{\bsX} (\bsx)$.

During the last few decades, various 
reliability analysis methods have been developed, and they can be categorized into four types: approximate analytical methods \cite{hasofer1974exact,der1987second}, numerical integration methods \cite{zhao2001moment,xu2019new}, surrogate-assisted methods \cite{echard2011ak,cheng2021adaptive,teixeira2021adaptive,moustapha2022active} and numerical simulation methods \cite{au2001estimation,papaioannou2016sequential,geyer2019cross,papaioannou2021combination}. The approximate analytical methods approximate the LSF by low-order Taylor expansions to estimate the failure probability. These include the first order reliability method (FORM) \cite{hohenbichler1982first} and the second order reliability analysis method (SORM) \cite{breitung1984asymptotic,kiureghian1991efficient}. However, the accuracy of these method can be low for highly non-linear problems. With numerical integration methods, the first few moments of the LSF are computed by the point estimation method \cite{zhao2001moment} or sparse grid integration method \cite{he2014sparse}, and the model response PDF as well as the the failure probability are estimated based on these moments. These methods suffer from the \textquotedblleft curse of dimensionality\textquotedblright, i.e., the associated computational cost rises sharply with the input variable dimension. Surrogate-assisted methods first fit a surrogate model based on a set of training points to approximate the true LSF, and then perform reliability analysis efficiently using the cheap-to-evaluate surrogate model. Recently, various surrogate model-based active learning strategies have been developed for reliability analysis \cite{echard2011ak,teixeira2021adaptive,moustapha2022active,dang2022structural,wei2023expected}. However, these methods only work for low-to-moderate  dimensional problems. Numerical simulation methods, based on Monte Carlo simulation (MCS), are generally preferred for reliability analysis of complex engineering systems due to their robustness.  In this category, various variance reduction methods have been developed, including: importance sampling (IS) \cite{au1999new,au2003important}, sequential importance sampling (SIS) \cite{papaioannou2016sequential,xian2024relaxation}, cross-entropy based adaptive IS algorithms \cite{geyer2019cross,papaioannou2019improved,kurtz2013cross,wang2016cross}, subset simulation (SuS) \cite{au2001estimation,papaioannou2015mcmc,wang2019hamiltonian}, line sampling (LS) \cite{papaioannou2021combination,pradlwarter2007application}, directional sampling (DS) \cite{ditlevsen1990general,bjerager1988probability}, directional importance sampling (DIS) \cite{grooteman2011adaptive,zhang2022cross}, sequential directional importance sampling (SDIS) \cite{cheng2023rare}.

In this paper, we present an enhanced SDIS method. In SDIS, the failure probability is decomposed into a group of integrals, defined by magnifying the input variability. The first integral is estimated with MCS, while the subsequent ones are estimated with directional importance sampling. 
SDIS has been shown to be efficient for problems with low magnitude of failure probability, multiple failure domains, and many input variables \cite{cheng2023rare}. However, the efficiency of SDIS depends on the initial magnification factor and the root-finding algorithm implemented to estimate the failure probability on every important direction. In this work, we enhance the performance of SDIS by the following modifications:

1. We note that the MCS estimator of the first integral in SDIS given in \cite{cheng2023rare} is biased, and provide an unbiased estimator. 

2. We discuss the efficiency of estimating the first integral with MCS in SDIS, and propose to estimate it with SuS. 

3. We propose a Kriging-based active learning algorithm for finding roots within a narrow confidence interval on every important direction. 

To evaluate the effect of these modifications, we compare the performance of the enhanced SDIS with the original SDIS and SuS on various benchmark reliability problems of different complexity. We find significant performance improvements for selected problems. 

The layout of this paper is as follows. In section 2, we review the background of MCS, IS, DS and DIS methods. In Section 3, we present the enhanced SDIS algorithm, where the SDIS estimator, resampling technique, MCMC algorithm, the Kriging-based active learning root-finding technique, and the SuS method for estimating the first integral in SDIS are presented. In Section 4, several benchmarks are used to compare the performance of SDIS and SuS. The paper concludes with final remarks in Section 5.

\section{Background}
\label{sec::Background}
 
As is common in structural reliability \cite{der_kiureghian_structural_2022}, the failure probability in Eq. \eqref{eq:Pf_x} is reformulated in standard normal space as 
\begin{equation}
P_f = \int_{\mathbb{R}^{n}} I\left(G(\bsu)\leq 0\right)\varphi_n (\bsu)  d\bsu=\mathbb{E}_{\varphi_n}\left[I\left(G(\bsU)\leq 0\right)\right] ,
\label{eq:Pf_u}
 \end{equation}
where $\varphi_n (\bsu)$ is the $n$-dimensional joint PDF of an independent standard normal random vector $\bsU$, and $G(\bsu)= g(\boldsymbol{T}^{-1}(\bsu))$ is the transformed LSF in the standard normal space, in which $\boldsymbol{T}$ is an isoprobabilistic transformation function \cite{hohenbichler1981non}.

MCS is the most straightforward and robust method to estimate the integral in Eq. \eqref{eq:Pf_u}. To implement it, one draws $N$ samples $\{\bsu_i,i=1,...,N\}$ from $\varphi_n (\bsu)$, evaluates $I\left(G(\bsu_i)\leq 0\right)$, and uses the sample mean to estimate $P_f$ as 
\begin{equation}
P_f \approx \hat P_f =  \frac{1}{N} \sum_{i=1}^N I\left(G(\bsu_i)\leq 0\right).
\label{eq:mcs}
 \end{equation}

 The MCS estimator in Eq. \eqref{eq:mcs} is unbiased, and its coefficient of variation (CoV) is given by
 \begin{equation}
  \delta =\sqrt{ \frac{1- P_f}{N P_f} }.
\label{eq:mcs_cov}
 \end{equation}
With $P_f$ in the order of $10^{-k}$, MCS approximately requires $10^{k+2}$ samples to obtain an estimator with $\delta=0.1$, which is infeasible for expensive computational models with low magnitude of failure probability. To reduce the computational cost, importance sampling (IS) reformulates the integral in Eq. \eqref{eq:Pf_u} as 
     \begin{equation}
      P_f = \int_{\mathbb{R}^{n}} I\left(G(\bsu)\leq 0\right)\frac{\varphi_n (\bsu)}{h(\bsu)} h(\bsu) d\bsu = \mathbb{E}_{h}\left[I\left(G(\bsU)\leq 0\right)W(\bsU)\right],
      \label{eq:isE}
      \end{equation}
where $h(\bsu)$ is the IS density, and $W(\bsu)=\varphi_n (\bsu)/{h(\bsu)}$ is the importance weight. In IS, one draws a group of samples $\{\bsu_i,i=1,...,N\}$ from $h(\bsu)$, and $P_f$ can be approximated as
\begin{equation}
P_f \approx \hat P_f =  \frac{1}{N} \sum_{i=1}^N I\left(G(\bsu_i)\leq 0\right)W(\bsu_i).
\label{eq:is}
 \end{equation}
The performance of IS depends on the choice of the IS density $h(\bsu)$, and the optimal IS density is given by
   \begin{equation}
       h_{opt}(\bsu)= \frac{I(G( \bsu)\leq 0) \varphi_n (\bsu)} {P_f}.
      \label{eq:IS_density}
    \end{equation}
 The estimator of Eq.~\eqref{eq:is} with IS density $h_{opt}(\bsu)$ has zero variance.
 However, $h_{opt}(\bsu)$ is not applicable in practice since it requires the knowledge of the failure probability $P_f$.
 
Alternatively, the failure probability can be estimated in polar coordinates through a coordinate transformation of the outcome space of the standard normal distribution.
 In polar coordinates, one can express the failure probability  as
     \begin{equation}
     \begin{aligned}
      P_f &=\int_{\mathbb{S}^{n-1}}\int_{0}^{\infty} I\left(G(r\bsa)\leq 0\right)f_{\chi_n}(r)f_A(\bsa) dr d\bsa = \mathbb{E}_{ f_{\chi_n},f_A}\left[I\left(G(R\bsA)\leq 0\right)\right] ,
      \label{eq:pf_polar}
       \end{aligned}
     \end{equation}
    where $\bsa$ is the direction and $r$ is the radius; $f_{\chi_n}(r)$ is the PDF of radius, given by the $\chi$ distribution with $n$ degrees-of-freedom, and $f_A(\bsa)$ is the PDF of direction on the $n-1$ dimensional unit hypersphere. In a similar manner to Eq.~\eqref{eq:isE}, the polar representation of the failure probability in Eq. \eqref{eq:pf_polar} can be reformulated by IS as 
       \begin{equation}
       P_f = \int_{\mathbb{S}^{n-1}}\int_{0}^{\infty} I\left(G(r\bsa)\leq 0\right)W(r,\bsa) h(r,\bsa) dr d\bsa = \mathbb{E}_{h}\left[I\left(G(R\bsA)\leq 0\right)W(R,\bsA)\right] ,
       \end{equation} 
where $h(r,\bsa)$ is the joint IS density of radius and direction, $W(r,\bsa)= f_{\chi_n}(r)f_A(\bsa)/{h(r,\bsa)}$ is the 
importance weight. The optimal zero-variance IS density in polar coordinates is given by
   \begin{equation}
       h_{opt}(r,\bsa)= \frac{I(G( r\bsa)\leq 0) f_{\chi_n}(r)f_A(\bsa)} {P_f}
      \label{eq:IS_density_polar}.
     \end{equation}
As with Eq.~\eqref{eq:IS_density}, the optimal IS density in Eq. \eqref{eq:IS_density_polar} cannot be implemented in practice. Alternatively, directional sampling (DS) makes use of the fact that one can reformulate the failure probability in Eq. \eqref{eq:pf_polar} as \cite{ditlevsen1990general}
    \begin{equation}
        \begin{aligned}
       P_f &=\int_{\mathbb{S}^{n-1}}\int_{0}^{\infty} I\left(G(r\bsa)\leq 0\right)f_{\chi_n}(r)dr f_A(\bsa) d\bsa   \\
       &=\int_{\mathbb{S}^{n-1}}{\rm Pr}\left(G(R\bsa)\leq 0\right)f_A(\bsa)d\bsa \\
       &= \mathbb{E}_{ f_A}\left[ {\rm Pr}\left(G(R\bsA)\leq 0\right)\right] \, ,
       \label{eq:ds}
        \end{aligned}
       \end{equation} 
 where the failure probability conditional on a specific direction $\bsa$ can be obtained by finding the roots of $G(r\bsa)=0$  \cite{zhang2022cross}, i.e., 
    \begin{equation}
       {\rm Pr}\left(G(R\bsa)\leq 0\right) 
    = \left \{
       \begin{aligned}
        \sum_{j=1}^{m}(-1)^j F_{\chi_n}\left(r_j(\bsa)\right),\,\ \qquad\qquad\quad\qquad\quad {\rm if} \,\ m \,\ \rm is \,\ even \\
       1-F_{\chi_n}\left(r_m(\bsa)\right) + \sum_{j=1}^{m-1}(-1)^j F_{\chi_n}\left(r_j(\bsa)\right), {\rm if}\,\ m \,\ \rm is \,\ odd
        \label{eq:conditional_probability}
        \end{aligned}
        \right.
       \end{equation} 
where $F_{\chi_n}(\cdot)$ is the CDF of $\chi$ distribution with $n$ degree-of-freedom, and $\{r_j(\bsa),j=1...,m\}$ are the roots of $G(r\bsa)= 0$ in direction $\bsa$. 

 To estimate $P_f$ with DS, one first draws a group of directional vectors $\{\bsa_i,i=1,...,N\}$ uniformly on the $n-1$ dimensional unit hyper-sphere, and then evaluates the conditional failure probabilities ${\rm Pr}\left(G(R\bsa_i)\leq 0\right)$ by finding the roots of $G(r\bsa_i)= 0$ in each direction $\bsa_i$. The DS estimator is finally given by 
\begin{equation}
P_f \approx \hat P_f =  \frac{1}{N} \sum_{i=1}^N {\rm Pr}\left(G(R\bsa_i)\leq 0\right).
\label{eq:ds_mcs}
 \end{equation}

 The above DS estimator is unbiased, but it is inefficient for high-dimensional problems since a large number of directions are required to sufficiently cover the $n-1$ dimensional unit sphere, and a large portion of them may point towards the safe domain. To alleviate this problem,  directional importance sampling (DIS) reformulates the DS estimator in Eq.  \eqref{eq:ds} as 
 \begin{equation}
       P_f=\int_{\mathbb{S}^{n-1}}{\rm Pr}\left(G(R\bsa)\leq 0\right) \frac {f_A(\bsa)}{h(\bsa)}h(\bsa)d\bsa = \mathbb{E}_{h}\left[{\rm Pr}\left(G(R\bsA)\leq 0\right) W(\bsA)\right],
      \label{eq:DS}
     \end{equation}
where $h(\bsa)$ is the DIS density, and $W(\bsa) = f_A(\bsa)/h(\bsa)$ is the directional importance weight. The optimal DIS density in theory is given by
     \begin{equation}
       h_{opt}(\bsa)= \frac{{\rm Pr}\left(G(R\bsa)\leq 0\right) f_A(\bsa)} {P_f}. 
      \label{eq:IS_density_DS}
    \end{equation}
As with the optimal IS densities in Eqs. \eqref{eq:IS_density} and \eqref{eq:IS_density_polar}, the density $ h_{opt}(\bsa)$ in Eq. \eqref{eq:IS_density_DS} cannot be used in practice. Recently, several adaptive methods have been proposed to design the DIS density \cite{grooteman2011adaptive,shayanfar2018adaptive,zhang2022cross}, but they are only effective for low-to-moderate dimensional problems.

\section{Enhanced sequential directional importance sampling  }

 In this section, we introduce the enhanced SDIS algorithm. To make this paper self-contained, we also recap the resampling technique, MCMC algorithm as well as the parameter choice strategy implemented in  \cite{cheng2023rare}.
 The novel elements of the enhanced SDIS compared to the SDIS method of \cite{cheng2023rare} are the unbiased MCS estimator of the first probability integral, discussed in Section~\ref{sec:SDIS}, the root-finding algorithm presented in Section~\ref{sec:rootfinding} and the SuS estimator of the first integral presented in Section~\ref{sec:Psigma1}.

\subsection{Sequential directional importance sampling }
\label{sec:SDIS}

 In SDIS, a family of auxiliary failure probabilities $\{P_{\sigma_1},\ldots,P_{\sigma_M}\}$ are defined by magnifying the variability of the input random variables as
\begin{equation}
\begin{aligned}
 P_{\sigma_i} &= \int_{\mathbb{R}^{n}} I\left(G(\sigma_i \bsu)\leq 0\right)\varphi_n (\bsu) d\bsu 
 &=\int_{\mathbb{S}^{n-1}} {\rm Pr}\left(G(\sigma_i R\bsa)\leq 0\right)f_A(\bsa)d\bsa ,
 \end{aligned}
 \end{equation}
where $\{\sigma_i,i=1,\ldots,M\},$ are magnification factors satisfying $1=\sigma_M \leq \cdots \leq \sigma_1$, and $G(\sigma_i \bsu)=G(\sigma_i r\bsa)$ is the $i$-th auxiliary LSF.

By presetting an initial magnification factor $\sigma_1$, the first auxiliary failure probability $P_{\sigma_1}$ in SDIS is estimated with MCS. In this step, one draws samples from $\varphi_n (\bsu)$ one-by-one until $n_s$ (with $n_s\in [100,200]$) samples in the inflated failure domain $F_{\sigma_1}=\{\bsu\in \mathbb{R}^{n} :G(\sigma_1 \bsu)\leq 0\}$  are generated. An estimate $\hat P_{\sigma_1}$ of $ P_{\sigma_1}$ is given by 
 \begin{equation}
 P_{\sigma_1} \approx \hat P_{\sigma_1} = \frac{n_s-1}{N-1} ,
\end{equation}
where $N$ is the total number of samples drawn from $\varphi_n (\bsu)$ until $n_s$ failure samples are obtained. This is the unbiased estimator of $P_{\sigma_1}$ in a Bernoulli process when a random number $N$ of samples are drawn until a fixed number of $n_s$ failure samples are obtained \cite{haldane1945method}. By contrast, the estimator $n_s/N$ used in \cite{cheng2023rare} is biased. An unbiased estimate of the variance of $\hat P_{\sigma_1}$ is given by \cite{finney1949method}
 \begin{equation}
  {\rm Var}[\hat P_{\sigma_1}] \approx   \frac{\hat P_{\sigma_1}(1-\hat P_{\sigma_1})}{N-2}.
\end{equation}
A corresponding estimate of the CoV of $\hat P_{\sigma_1}$ is
 \begin{equation}
  \delta_{\hat P_{\sigma_1}} \approx  \sqrt{\frac{1-\hat P_{\sigma_1}}{(N-2)\hat P_{\sigma_1}}}.
\end{equation}
 
In SDIS, the subsequent auxiliary failure probabilities $P_{\sigma_i}(i>1)$ are expressed by DIS as 
  \begin{equation}
  \begin{aligned}
 P_{\sigma_i} &= \int_{\mathbb{R}^{n}} I\left(G(\sigma_i \bsu)\leq 0\right)\varphi_n (\bsu) d\bsu 
\\ &=\int_{\mathbb{S}^{n-1}} {\rm Pr}\left(G(\sigma_i R\bsa)\leq 0\right)f_A(\bsa)d\bsa
 \\ &= P_{\sigma_{i-1}}\int_{\mathbb{S}^{n-1}}\frac{ {\rm Pr}\left(G(\sigma_i R\bsa)\leq 0\right)}{ {\rm Pr}\left(G(\sigma_{i-1} R\bsa)\leq 0\right)}\frac{{\rm Pr}\left(G(\sigma_{i-1} R\bsa)\leq 0\right)f_A(\bsa)}{P_{\sigma_{i-1}}}d\bsa\\ &=P_{\sigma_{i-1}} \mathbb{E}_{h_{\sigma_{i-1}}}\left[\frac{ {\rm Pr}\left(G(\sigma_i R\bsa)\leq 0\right)}{ {\rm Pr}\left(G(\sigma_{i-1} R\bsa)\leq 0\right)}\right]\\ &=P_{\sigma_{i-1}} \mathbb{E}_{h_{\sigma_{i-1}}}\left[W_{i-1}(\boldsymbol{A})\right] ,
\end{aligned}
\end{equation}
where  
\begin{equation}
h_{\sigma_{i-1}}(\bsa)=\frac{{\rm Pr}\left(G(\sigma_{i-1} R\bsa)\leq 0\right)f_A(\bsa)}{P_{\sigma_{i-1}}}
\end{equation}
is the optimal DIS density of the $i-1$-th auxiliary reliability problem, and   
\begin{equation}
W_{i-1}(\bsa) =\frac{{\rm Pr}\left(G(\sigma_{i} R\bsa)\leq 0\right)}{{ {\rm Pr}\left(G(\sigma_{i-1} R\bsa)\leq 0\right)}}.
\label{eq:directional_weight}
\end{equation}
is the directional importance weight. 
%
%
%
In this manner, the failure probability $P_f$ is expressed as
\begin{equation}
 P_f =P_{\sigma_1}\prod_{i=1}^{M-1} S_i,
\end{equation}
where $S_i=\mathbb{E}_{h_{\sigma_i}}\left[W_i(\boldsymbol{A})\right] $ is the ratio between two adjacent auxiliary failure probabilities. It can be estimated by the sample mean of the directional importance weight $W_i(\boldsymbol{A})$ as 
\begin{equation}
S_i \approx \hat S_i = \frac{1}{n_s}\sum_{k=1}^{n_s}\frac{ {\rm Pr}\left(G(\sigma_{i+1} R\bsa_k)\leq 0\right)}{ {\rm Pr}\left(G(\sigma_{i} R\bsa_k)\leq 0\right)} ,
\label{eq:weight_mean}
\end{equation}
where $\{\bsa_k,k=1,...,n_s\}$ are drawn from the DIS density $h_{\sigma_i}(\bsa)$. 
The magnification factors $\sigma_i,i>1,$ are determined on the fly as detailed in Section \ref{sec:parSDIS}.

 Both ${\rm Pr}\left(G(\sigma_{i+1} R\bsa)\leq 0\right)$ and ${\rm Pr}\left(G(\sigma_{i} R\bsa)\leq 0\right)$ in Eq. \eqref{eq:weight_mean} can be estimated following Eq. \eqref{eq:conditional_probability}, i.e., by finding the roots of $G(\sigma_{i+1} r\bsa)= 0$ and $G(\sigma_{i} r\bsa)= 0$  in direction $\bsa$. In practice, one only needs to find the roots $\{r_{i+1,j}(\bsa),j=1,...,m\}$ of $G(\sigma_{i+1} r\bsa)= 0$; the roots of $G(\sigma_{i} r\bsa)= 0$ in the same direction $\bsa$ can be obtained by forcing $\sigma_{i+1} r_{i+1,j}(\bsa)\bsa = \sigma_{i} r_{i,j}(\bsa)\bsa $ for $j=1,...,m$, as
\begin{equation}
r_{i,j}(\bsa) = \frac{\sigma_{i+1}r_{i+1,j}(\bsa)}{\sigma_{i}}, j=1,...,m.
\label{eq:root_relationship}
\end{equation}

\subsection{Resampling  algorithm}
\label{sec:resampling}
In the SDIS procedure, one needs to sample a directional vector $\bsa$ from the DIS density function $h_{\sigma_i}(\bsa)= {\rm Pr}\left(G(\sigma_{i} R\bsa)\leq 0\right)f_A(\bsa)/{P_{\sigma_{i}}} $. 
To draw directional vectors $\bsa$ from the first DIS density $h_{\sigma_1}(\bsa)$, one can apply a polar coordinate transformation to the $n_s$ failure samples $\{\bsu_i,i=1,...,n_s\}$ following 
 $h_{\sigma_1}(\bsu)= I\left(G(\sigma_{1} \bsu)\leq 0\right)\varphi_n(\bsu)/{P_{\sigma_{1}}}$ obtained in the first step when estimating $P_{\sigma_{1}}$ with MCS.
 %
 Indeed, for a sample $\bsu$ drawn from $h_{\sigma_i}(\bsu)$ in Cartesian coordinates, the corresponding polar representation  $\{\bsa,r\}$ follow the following joint IS density due to transformation of variables: 
  \begin{equation}
    h_{\sigma_i}(r,\bsa) =
    \frac{I(G( \sigma_ir\bsa)\leq 0) f_{\chi_n}(r)f_A(\bsa)} {P_{\sigma_{i}}}.
  \label{IS_density_polar}
  \end{equation} 
  The DIS density $h_{\sigma_i}(\bsa)$ is simply the marginal distribution of $\bsa$, namely
  \begin{equation}
    h_{\sigma_i}(\bsa) = \int_0^\infty h_{\sigma_i}(r,\bsa)dr= \frac{{\rm Pr}\left(G(\sigma_{i} R\bsa)\leq 0\right)f_A(\bsa)}{P_{\sigma_{i}}}.
 \label{DIS_density}
\end{equation} 
As a result, by normalizing samples $\bsu$ drawn from IS density $h_{\sigma_i}(\bsu)$ in Cartesian coordinates, one obtains the corresponding directional vector $\bsa = \bsu/\Vert\bsu\Vert $ that follows the DIS density $h_{\sigma_i}(\bsa)$ in polar coordinates.

In the subsequent steps, a resample-move scheme is applied to draw samples from $h_{\sigma_i}(\bsa)(i>1)$. One first resamples a group of initial samples from existing samples obtained from the previous sampling step and then moves these samples to regions of high probability mass of $h_{\sigma_i}(\bsa)$ by applying MCMC sampling algorithm with invariant distribution $ h_{\sigma_i}(\bsa)$. However,  evaluation of the acceptance probability in classical Metropolis-Hastings (M-H) algorithms, as detailed in Section \ref{sec:mcmc}, requires evaluation of the target distribution $h_{\sigma_i}(\bsa)$ and, hence, the conditional failure probability ${\rm Pr}\left(G(\sigma_{i} R\bsa)\leq 0\right)$, at the candidate state.
Each evaluation of this probability requires finding the roots of $G(\sigma_{i} r\bsa)=0$.
To avoid this problem, we suggest to apply a resample-move scheme for drawing samples $\{\bsu_k,k=1,...,n_s\}$ from the distribution $h_{\sigma_i}(\bsu)= I\left(G(\sigma_{i} \bsu)\leq 0\right)\varphi_n(\bsu)/{P_{\sigma_{i}}}$. The directional vector following $h_{\sigma_i}(\bsa)$ is then obtained by normalizing the samples drawn from $h_{\sigma_i}(\bsu)$, namely, $\{\bsa_k=\bsu_k/\Vert\bsu_k\Vert,k=1,...,n_s\}$.

 To draw initial samples from $h_{\sigma_i}(\bsu)$, one can apply the following resampling algorithm:

1. Resampling of direction: select $n_a$ samples $\{\bsa_k,k=1,...,n_s\}$ with replacement from the existing directional samples following $h_{\sigma_{i-1}}(\bsa)$ randomly with probability assigned to each sample proportional to the importance weight $W_{i-1}(\bsu)$.

2. Resampling of radius: For each directional vector $\bsa_k(k=1,...,n_s)$ obtained in step 1, generate a random sample of the radius $r(\bsa_k)$ following the truncated $\chi_n$ distribution $f_{\chi_n}(r|F_{\bsa_k})$, where $F_{\bsa_k}=\{r\in \mathbb{R} :G(\sigma_i r\bsa_k)\leq 0\}$ is the failure domain in direction $\bsa_k$, given by  
\begin{equation*}
 F_{\bsa_k} = \left \{
 \begin{aligned} r\in \bigcup_{j\in \{1,3,...,m-1\}}[r_{i,j}(\bsa_k), r_{i,j+1}(\bsa_k)], \,\ \qquad\qquad {\rm if} \,\ m \,\ {\rm is \,\ even}\\
  r\in \bigcup_{j\in \{1,3,...,m-2\}}[r_{i,j}(\bsa_k), r_{i,j+1}(\bsa_k)]\cup [r_{i,m}(\bsa_k),+\infty], {\rm if} \,\ m \,\ {\rm is \,\ odd}
 \end{aligned}
\right.
\end{equation*}

The resampled sample pairs $\{\bsa_k,r(\bsa_k),k=1,...,n_s\}$ follow the joint IS density function $ h_{\sigma_i}(\bsa,r)$ asymptotically, and hence, the corresponding samples in Cartesian coordinates  $\{\bsu_k=\bsa_kr(\bsa_k),k=1,...,n_s\}$ follow $ h_{\sigma_i}(\bsu)$ asymptotically. 

\subsection{
MCMC algorithm}
\label{sec:mcmc}

Following the resampling step, MCMC sampling is applied to generate directional samples following $h_{\sigma_i}(\bsa)$.
As discussed in Section \ref{sec:resampling}, this is achieved through first generating samples $\{\bsu_k,k=1,...,n_s\}$ from $h_{\sigma_i}(\bsu)$ using as seeds the Cartesian samples obtained in the resampling step, and then transforming the generated samples to directional samples, i.e., applying $\{\bsa_k=\bsu_k/\Vert\bsu_k\Vert,k=1,...,n_s\}$.

For each seed $\{\bsu_k,k=1,...,n_s\}$ obtained in the resampling step, one simulates a Markov chain of length $l$ independently with stationary distribution  $h_{\sigma_i}(\bsu)$  based on the M-H algorithm as follows:
   
   1. Select a seed $\bsu_t$, and set $t=0$.

2. Generate a random candidate sample pair  $\bsu'$ from the proposal distribution $q_{\sigma_i}(\bsu'|\bsu_t)$, and evaluate the model response $G(\sigma_i\bsu')$.

   3. Calculate the acceptance probability
 \begin{equation}
  \alpha = I(G(\sigma_i\bsu')\leq 0){\rm min}\left(1, \frac{\varphi_n(\bsu') q(\bsu_t|\bsu')} {{\varphi_n(\bsu_t) q(\bsu'|\bsu_t)}}\right) .
 \label{eq:acceptance_rate}
\end{equation}

4. Generate a random number $w\in [0,1]$, and determine the next state pair as
\begin{equation}
  \bsu_{t+1} =\begin{cases}
 \bsu_t \quad\quad \,\ \,\ {\rm if} \,\ w\geqslant \alpha \\  \bsu' \quad\qquad  {\rm if}  \,\ w< \alpha 
  \end{cases}; 
\end{equation}

5. Set $t = t + 1$, and go back to step 2 until the target length $l=5$ is reached.

In SDIS, the conditional sampling (CS) method \cite{papaioannou2015mcmc} is applied, in which the proposal distribution $q_{\sigma_i}(\bsu'|\bsu_t)$ is selected as the multivariate standard Gaussian density conditional on the current state $\bsu_t$
  \begin{equation}
   \bsu'= \rho_{\bsu} \bsu_t + \sqrt{1-\rho_{\bsu}^2}\boldsymbol{\xi} ,
 \label{eq:direction}
 \end{equation}
where  $\boldsymbol{\xi}\in \mathbb{R}^{n}$ is a realization of an $n$-dimensional independent standard normal random vector, and $\rho_{\bsu}$ is the correlation coefficient between $\bsu_t$ and $\bsu'$. The correlation parameter $\rho_{\bsu}$ is determined 
adaptively to ensure that the acceptance probability $\alpha$ is close to a near-optimal value $44\%$ \cite{papaioannou2015mcmc}. 

With the above proposal, the MCMC acceptance probability in Eq. \eqref{eq:acceptance_rate} becomes
\begin{equation}
 \alpha = I(G(\sigma_i\bsu')\leq 0).
\end{equation}

  To reduce the effect of the transient period of the simulated Markov chain and reduce the correlation between Markov Chain states, we discard the first $l$ samples from every Markov chain. This means that only the last state of each Markov chain is retained, which is expected to follow the target distribution $h_{\sigma_i}(\bsu)$ closely.

\subsection{Choice of parameters for SDIS}
\label{sec:parSDIS}
 There are several user-specified parameters in SDIS, including the initial magnification factor $\sigma_1$, the number of important directional vectors per level $n_s$, and the Markov chain length $l$. Based on our previous work \cite{cheng2023rare} and numerical experiments, we suggest setting $\sigma_1=3, n_s \in [100,200], l=5$. 

The magnification factor in the subsequent steps $\{\sigma_i,i>1\}$ can be determined on the fly by forcing the CoV of the directional importance weight to  adhere to a target value $\delta_{\rm target}$, which corresponds to solving the following optimization problem \cite{papaioannou2016sequential}:
\begin{equation}
 \sigma_i = \underset{\sigma_i \in [1, \sigma_ {i-1}] }{\mathrm{argmin}}|\hat\delta_{W_i}-\delta_{\rm target}|, (i \geqslant 2),
 \label{eq:opt_sigma}
\end{equation}
where $\hat\delta_{W_i}$ is the CoV of the importance weight $\{W_i(\bsa_j),j=1,...,n_s\}$. We use $\delta_{\rm target}=1.5$ for the target CoV. 

Assuming that directional vectors drawn from $h_{\sigma_i}(\bsa)$ are independent from samples drawn at other steps $h_{\sigma_j}(\bsa)(i\neq j)$, and that the directional vectors at each level drawn from $h_{\sigma_i}(\bsa)$ are independent from each other, the CoV of the SDIS estimator can be approximated by \cite{cheng2023rare}
\begin{equation}
 \delta_{\hat P_f} \approx \sqrt{\delta_{\hat P_{\sigma_1}}^2 + \sum_{i=1}^{M-1}\delta_{\hat S_i}^2},
 \label{eq:cov_sdis}
\end{equation}
where $\delta_{\hat 
 P_{\sigma_1}}$ is the CoV of $\hat 
 P_{\sigma_1}$, and $\delta_{\hat S_i}$ is the CoV of $S_i$, which is given by
 \begin{equation}
  \delta_{\hat S_i} \approx \frac{\delta_{ W_i(\boldsymbol{A})}}{\sqrt{n_s}}. 
 \end{equation}
 $\delta_{ W_i(\boldsymbol{A})}$ denotes the CoV of the importance weight $W_i(\boldsymbol{A})$, and it is estimated from the existing samples $\{W_i(\bsa_i),i=1,...,n_s\}$. Note that $\delta_{\hat P_f}$ underestimates the true CoV of $\hat P_f$ due to the independence assumption.

\subsection{Root-finding algorithm}
\label{sec:rootfinding}

 To estimate the conditional probability ${\rm Pr}\left(G(\sigma_i R\bsa)\leq 0\right)$ in Eq. \eqref{eq:conditional_probability}, one needs to find the roots of $G(\sigma_ir\bsa)=0$ along direction $\bsa$. In \cite{cheng2023rare}, the 
trust-region dogleg algorithm (“fsolve” function in Matlab)
 is applied. However, it can only find one root along each direction. To address this issue, we introduce a Kriging-based active learning algorithm for identifying the roots in this section. 

 Since the radius $r$ follows the ${\chi_n}$ distribution, we limit the search domain on the radius to the following  interval 
 \begin{equation}
 \left[r^-,r^+\right]= \left[{F^{-1}_{\chi_n}(\alpha/2)},  F^{-1}_{\chi_n}(1-\alpha/2) \right],
\label{eq:confidence_interval}
 \end{equation}
 where $\alpha$ is a small positive constant.  As $(1-\alpha)\%$ probability mass of the PDF of radius concentrates on this interval, we can obtain an accurate estimation of the failure probability ${\rm Pr}\left(G(\sigma_i R\bsa)\leq 0\right)$ by finding the roots of $G(\sigma_ir\bsa)=0$ within this interval in direction $\bsa$. In Table \ref{tab:confidence_interval}, a series of intervals are listed by varying the input dimension $n$ for $\alpha=10^{-10}$. Although the lower bound and upper bound increase with $n$, it is observed that the interval width converges to a constant as $n\rightarrow\infty $.
 This is due to the fact that the ${\chi_n}$ distribution is approximately normal as $n\rightarrow\infty $ with variance $1 / 2$, which is independent of $n$ \cite{johnson1994continuous}. Therefore, to find the roots along each direction we only need to construct an accurate Kriging model within a narrow interval.  In SDIS, the roots of $G(\sigma_ir\bsa)=0$ will be used to compute the roots of $G(\sigma_{i+1}r\bsa)=0$ via the relationship in Eq. \eqref{eq:root_relationship}. 
   The roots of the first few auxiliary LSFs $G(\sigma_ir\bsa)=0(i=1,...)$ 
    may be smaller than the lower bound $F^{-1}_{\chi_n}(\alpha/2)$ given in Eq. \eqref{eq:confidence_interval}. To ensure that all roots of $G(\sigma_ir\bsa)=0$ are contained in the interval, we apply a very conservative lower bound $r^-=F^{-1}_{\chi_n}(\alpha/2)/\sigma_i$ in practice.

\begin{table}[hbt!]
\setlength{\belowcaptionskip}{1pt}
\caption{Interval defining the search domain on the radius for  $\alpha=10^{-10}$ and different dimension $n$ }
\centering
  \begin{tabular}{cccc}
  \hline
  Dimension $n$ &  Lower bound $r^-$ &  Upper bound $r^+$  & Width \\
  \hline
  10 &  $0.21$ &  $8.35$ &  $8.14$ \\
  $10^2$ &  $5.80$ &  $14.84$ &  $9.04$\\
  $10^3$ &  $27.16$ &  $36.29$&  $9.13$\\
  $10^4$ &  $95.46$ &  $104.60$&  $9.14$\\
   $10^5$ &  $311.67$ &  $320.81$&  $9.15$\\
    $10^6$ &  $995.43$ &  $1004.57$&  $9.15$\\
 \hline
\end{tabular}
\label{tab:confidence_interval}
\end{table}

Given an initial training set $\{\boldsymbol{X},\boldsymbol{Y}\}$ on direction $\bsa$, where $\boldsymbol{X}=[r^{(1)},...,r^{(M)}]^{\rm T}$, $\boldsymbol{Y}=[G(\sigma_i r^{(1)}\bsa ),...,G(\sigma_i r^{(M)}\bsa )]^{\rm T}$, we construct a 1-dimensional Kriging model as \cite{kleijnen2009kriging} 
\begin{equation}
\hat Y(r) \sim \mathcal {N}\left(\mu(r), s^2(r)\right) ,
\label{eq::kriging}
\end{equation}
where  $\mu(r)$ and $s^2(r) $ are the Kriging mean and variance respectively. In the current work, the Matérn 5/2 correlation function is used.  More details of the Kriging model are provided in  \ref{sec:Kriging}.
  
 We then select informative samples to refine the Kriging model
 sequentially. We apply the learning function developed in \cite{yang2015active} to select the optimal next point, which is defined as follows: 
\begin{equation}
   \begin{aligned}
  {\rm L}(r) &= \begin{cases}
     \mathbb{E}[{\rm max}(\hat Y(r),0)], \,\  \qquad  \qquad \quad{\rm if} \,\ \mu(r) < 0 \\
  \mathbb{E}[{\rm max}(-\hat Y(r),0)], \qquad\quad \,\ \,\ \,\ {\rm if} \,\  \mu(r) > 0  
    \end{cases} 
\end{aligned}
 \end{equation}
 It measures the average  distance of the predicted response to the limit state surface $G(\sigma_i r\bsa)=0$ when it is misclassified. The analytical expression of this function is given by \cite{yang2015active} 
\begin{equation}
   \begin{aligned}
    {\rm L}(r) &= -{\rm sgn}\left(\mu(r)\right)\mu(r)\Phi\left(-{\rm sgn}(\mu(r))\frac{\mu(r)}{s(r)}\right) +s(r) \varphi\left(\frac{\mu(r)}{s(r)}\right) ,
\end{aligned}
 \end{equation}
where $\Phi(\cdot)$ and $\varphi(\cdot)$ are the CDF and PDF of the standard normal distribution respectively, and ${\rm sgn}(\cdot)$ is the sign function. The convergence criterion used in this work is ${\rm max}({\rm L}(r))/\bar{|y|}
 <\epsilon$, where ${\rm max}({\rm L}(r))$ is the maximum of ${\rm L}(r)$, $\bar {|y|} $ is the mean of the absolute value of the model response in the training set and $\epsilon$ is a user-specified constant. A small $\epsilon$ will yield more accurate roots but requires more model evaluations, and vice versa. Here we set $\epsilon =5\times 10^{-4}$.

In the current work, we construct a Kriging model with a few initial radius samples in every direction. The initial sample size (3 or 4) is determined heuristically, depending on the degree of nonlinearity of the problem at hand. The first radius sample is the origin $r^{(1)}=0$, which is shared by all directions; the second sample is the radius of the failure sample $\bsu$ that defines the current direction, namely, $r^{(2)}=\Vert\bsu\Vert$; the third sample is determined as
\begin{equation}
r^{(3)}= \begin{cases}
(r^-+r^{(2)})/2, \,\ {\rm if} \,\ |r^{(2)}-r^+| < |r^+ - r^-|/3, \\
(r^+ + r^{(2)})/2, \,\  {\rm else}.
\end{cases}
\label{eq:third_radius}
\end{equation}
 
With the setting in Eq. \eqref{eq:third_radius}, we  put the third radius sample at the midpoint between the second one and either the lower bound $r^-$ or the upper bound, depending on the distance $|r^{(2)}-r^+|$. If the average sample size of the Kriging model after active learning along previous $n_{s1}$ directions is greater than 4, we turn to train the Kriging model along the remaining $n_{s}-n_{s1}$ directions with 4 initial radius samples. The location of the fourth radius sample is set to
\begin{equation}
r^{(4)}= \begin{cases}
(r^{(3)}+r^{(2)})/2, \,\ {\rm if} \,\ |r^{(2)}-r^+| < |r^+ - r^-|/3 \,\ {\rm and} \,\  G(\sigma_i r^{(3)}\bsa) > 0, \\
(r^- +r^{(3)})/2, \,\ \,\   {\rm if} \,\ |r^{(2)}-r^+| < |r^+ - r^-|/3 \,\ {\rm and} \,\ G(\sigma_i r^{(3)}\bsa) < 0, \\
(r^- +r^{(2)})/2, \,\ \,\   {\rm else}
\end{cases}
\label{eq:fouth_radius}
\end{equation}

\begin{figure}[htp]
  \centering \includegraphics[width=1\textwidth,trim={20 285 20 290},clip]{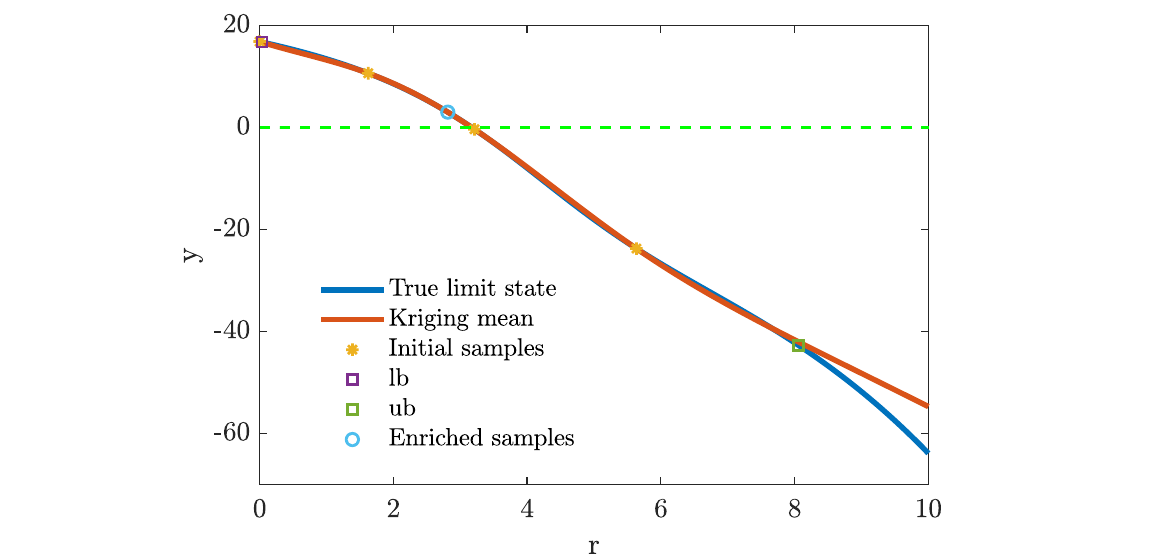}
  \caption{Active learning Kriging model for finding root (one root case) in interval $[0.0362,8.0574]$. }
  \label{fig:Kriging_root1}
  \end{figure}

\begin{figure}[htp]
  \centering \includegraphics[width=1\textwidth,trim={20 285 20 300},clip]{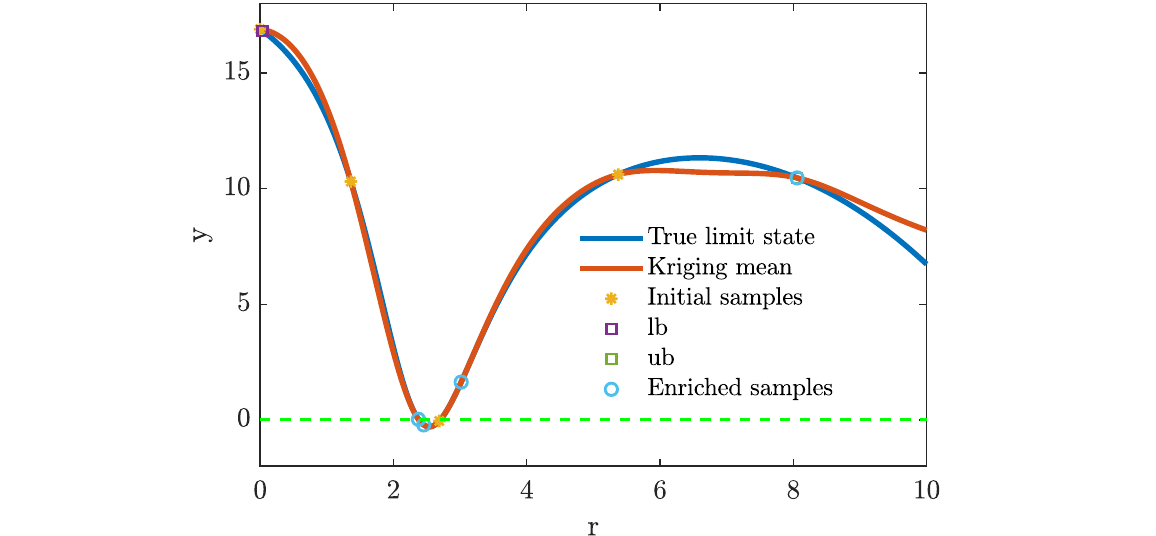}
  \caption{ Active learning Kriging model for finding roots (two roots case) in interval $[0.0362,8.0574]$.  }
  \label{fig:Kriging_root2}
  \end{figure}

\begin{figure}[htp]
  \centering \includegraphics[width=1\textwidth,trim={20 285 20 290},clip]{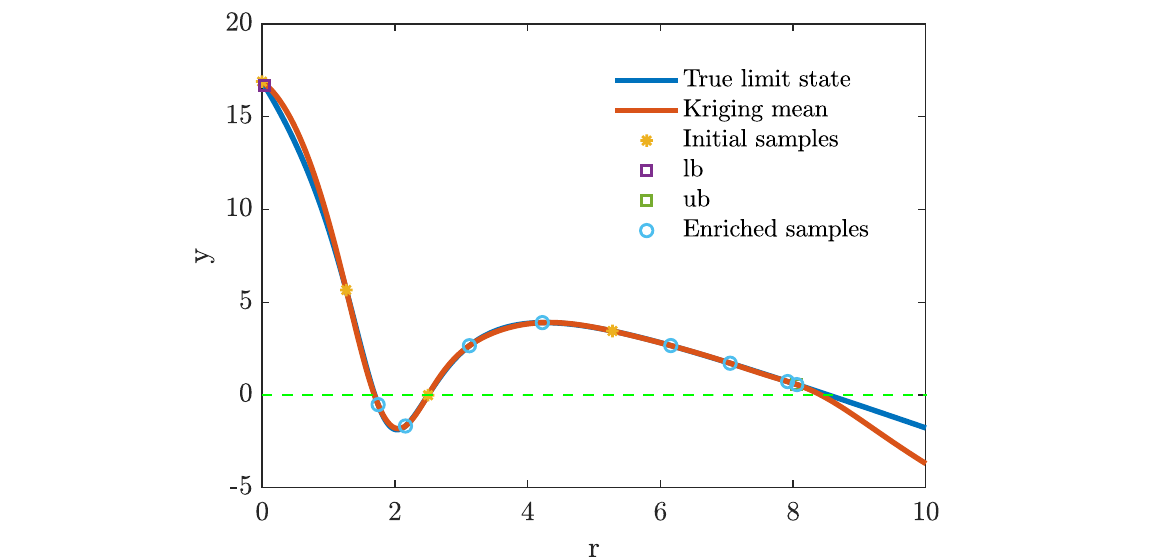}
  \caption{Active learning Kriging model for finding roots (three roots case) in interval $[0.0362,8.0574]$.  }
  \label{fig:Kriging_root3}
  \end{figure}

In Figs. \ref{fig:Kriging_root1} ,\ref{fig:Kriging_root2} and \ref{fig:Kriging_root3}, we present three examples of the active learning Kriging model for finding the root of $G(\sigma_1r\bsa)=0$ in the Example of Section \ref{example:nonlinear_oscillator} in the interval $[0.0362,8.0574]$. 
The figures show that the Kriging model can find all roots in these cases, and the enriched samples are quite close to the failure boundary, thereby refining the accuracy of the roots. 

\subsection{The efficiency of estimating $P_{\sigma_1}$ }
\label{sec:Psigma1}

  \begin{figure}[htp]
  \centering
\includegraphics[width=1\textwidth,trim={0 315 0 315},clip]{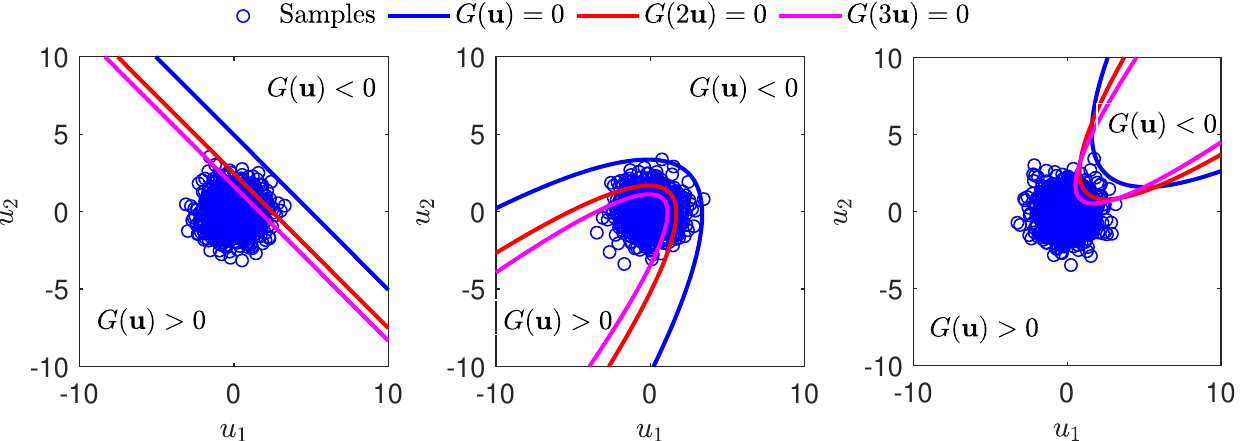}
  \caption{1000 MCS samples drawn from standard normal PDF $\varphi_2(\bsu)$ vs three types of limit states (left: linear; middle: concave; right: convex)}
  \label{fig:p1}
  \end{figure} 

In SDIS, the efficiency of estimating $P_{\sigma_1}$ with MCS in the first step depends on the initial magnification factor $\sigma_1$, which is set to 3 in \cite{cheng2023rare,rashki2021sesc} empirically. For most problems, this setting is efficient for generating failure samples lying in $F_{\sigma_1}$. To further illustrate its effectiveness, we consider three different types of limit state surfaces in 2-dimensional space, depicted in Fig. \ref{fig:p1}. For all the three cases, i.e., linear, concave and convex, the corresponding auxiliary limit state surfaces $G(\sigma\bsu)$ are much closer to the origin compared to the true ones, and one can obtain a group of failure samples lying in  $F_{\sigma_1}=\{\bsu\in \mathbb{R}^{n} :G(\sigma_1 \bsu)\leq 0\}$ efficiently with MCS by drawing samples from $\varphi_2(\bsu)$. 

\begin{figure}[htp]
  \centering
\includegraphics[width=1\textwidth,trim={10 300 10 300},clip]{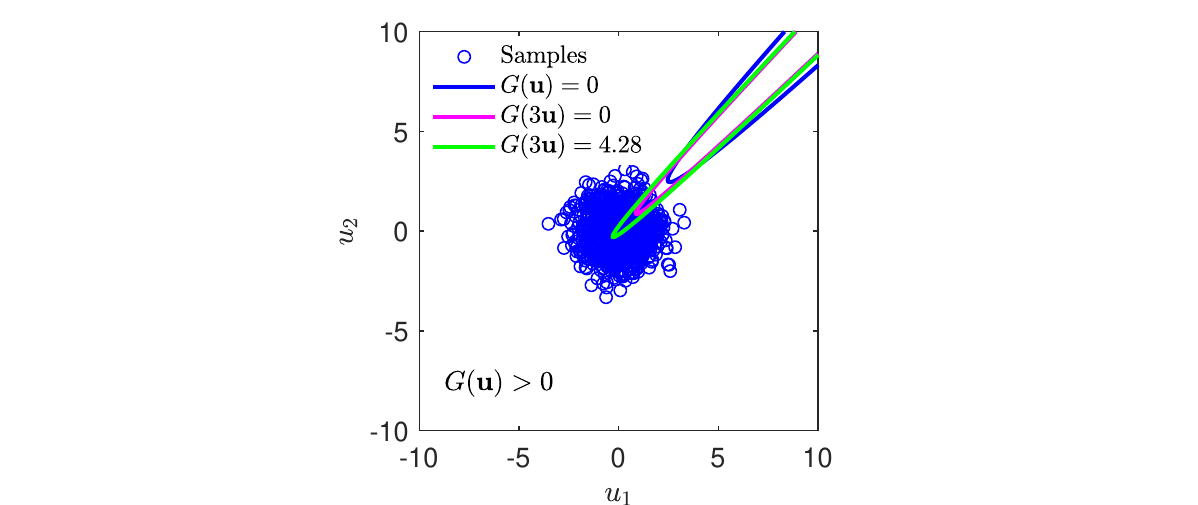}
  \caption{ 1000 samples drawn from standard normal PDF vs a special LSF with narrow failure domain. The probability of a sample to fall into the inflated failure domain $F=\{\bsu\in \mathbb{R}^{n} :G(3\bsu)\leq 0\}$ is 0.014. }
  \label{fig:p1_fail}
  \end{figure}

However, for problems with extremely low magnitude of failure probability or with a very narrow (e.g., island-shape) failure domain, only a small portion of the MCS population will reside in the failure domain $F_{\sigma_1}$. In Fig. \ref{fig:p1_fail}, we present such an example, in which case MCS is inefficient for estimating $P_{\sigma_1}$. To address this problem, we suggest introducing another family of intermediate failure events $G(\sigma_1 \bsu)\leq b_j$ by modifying the failure threshold \cite{cheng2022estimation,xian2024relaxation}. In other words, we estimate $P_{\sigma_1}$ with SuS. When $10n_s$ samples have been generated from $\varphi_n(\bsu)$ one-by-one before $n_s$ failure samples are obtained, we turn to estimate $P_{\sigma_1}$ with SuS, which gives 
\begin{equation}
 P_{\sigma_1} = \prod_{j=1}^s {\rm Pr}(\bsU \in F_{\sigma_1,b_j}| \bsU \in F_{\sigma_1,b_{j-1}}),
 \label{eq:sus_estimator}
\end{equation}
where $F_{\sigma_1,b_j}=\{\bsu\in \mathbb{R}^{n} :G(\sigma_1 \bsu)\leq b_j\}$ are intermediate failure domains, and $\infty=b_0>b_1> \cdots >b_s=0$ are the corresponding intermediate failure thresholds. The thresholds $b_j(j=1,...,s)$ are set as the $p_0$-percentiles (we set $p_0=0.1$ in this study) of the limit-state function values $\{G(\sigma_1 \bsu_i),i=1,...,n_s\}$, in which $\bsu_i$ are sampled from $\varphi_n (\bsu)$ in the initial step $j=1$ with crude Monte Carlo, and from the IS density $h_{\sigma_1,b_j}(\bsu) \propto I\left(G(\sigma_1 \bsu)\leq b_j\right)\varphi_n (\bsu)$ in steps $j = 2,...,s$ by MCMC through application of the adaptive conditional sampling algorithm \cite{papaioannou2015mcmc}.  In the last step of SuS, we obtain a group of failure samples in $F_{\sigma_1}=\{\bsu\in \mathbb{R}^{n} :G(\sigma_1 \bsu)\leq 0\}$ drawn from the last IS density  $h_{\sigma_1}(\bsu) \propto I\left(G(\sigma_1 \bsu)\leq 0\right)\varphi_n (\bsu)$. By normalizing these samples, i.e., by applying $\bsa = \bsu/\Vert\bsu\Vert$, one can obtain a group of directional vectors following the first DIS density $h_{\sigma_1}(\bsa)$. 
An estimate of the CoV of the SuS estimator of $P_{\sigma_1}$ is given in \cite{au2001estimation}.

\begin{remark}
    The SuS estimator $P_{\sigma_1}$ in Eq. \eqref{eq:sus_estimator} is biased due to the correlation between the estimates of the conditional probabilities ${\rm Pr}(\bsU \in F_{\sigma_1,b_j}|\bsU \in F_{\sigma_1,b_{j-1}})$ as well as the adaptive estimation of the intermediate failure domains  $F_{\sigma_1,b_j}=\{\bsu\in \mathbb{R}^{n} :G(\sigma_1 \bsu)\leq b_j\}$ \cite{papaioannou2015mcmc,cerou2012sequential}. However, this bias is small compared to the CoV of the estimator, hence it is not deemed critical.
\end{remark}

\subsection{Summary of enhanced SDIS}

The enhanced SDIS algorithm is summarized as follows:

1. Initialize: Set the initial magnification factor $\sigma_1$ = 3, the length of each Markov chain $l = 5$ and
the number of the important directions per level $n_s (n_s\in [100,200])$. Set $k=1$. 

2. Estimate the first auxiliary failure probability $\hat P_{\sigma_1}$ with MCS or SuS and determine the directional samples following $h_{\sigma_{1}}(\bsa)$ in terms of the failure samples $\{\bsu_i,i=1,...,n_s\}$ through applying $\{\bsa_i = \bsu_i/\Vert\bsu_i\Vert,i=1,...,n_s\}$.

3. Find the roots in the $n_s$ important directions $\{\bsa_i ,i=1,...,n_s\}$ with active learning Kriging model.

4. Determine the next magnification factor $\sigma_{k+1}$ by solving the optimization problem in Eq. \eqref{eq:opt_sigma}.

5. Calculate the ratio $\hat S_k$ between two adjacent auxiliary failure probabilities $P_{\sigma_{k+1}}$ and $P_{\sigma_k}$  with Eq. \eqref{eq:weight_mean}

6. Resample and move: Apply the resample and move algorithm introduced in Sections \ref{sec:resampling} and \ref{sec:mcmc} to draw a group of samples $\{\bsa_i,i=1,...,n_s\}$ 
following $h_{\sigma_{k+1}}(\bsa)$.

7. Set $k=k+1$, and go back to step 3 until $\sigma_{k+1} =1$.

8. Calculate the final failure probability as
\begin{equation}
\hat P_f =\hat P_{\sigma_1}\prod_{i=1}^{k} \hat S_i.
\end{equation}

\section{Numerical examples}
\label{sec:numerical_example}
In this section, the effectiveness of the enhanced SDIS is investigated on several benchmarks. We compare it to SuS \cite{au1999new,papaioannou2015mcmc} and standard SDIS \cite{cheng2023rare} in terms of the relative efficiency with respect to crude MCS, a concept that is borrowed from statistics \cite{l1994efficiency,chan2023bayesian}. This relative efficiency measure is defined as follows:
 \begin{equation}
   {\rm relEff}(\hat P_f) := \frac{P_f(1-P_f)}{{\rm MSE}(\hat P_f){\rm Cost}(\hat P_f)} .
   \label{eq:releff}
 \end{equation}
$P_f$ is the failure probability reference value obtained with MCS or analytically. The mean square error  ${\rm MSE}(\hat P_f)= (P_f- \mathbb{E}[\hat P_f])^2+ \mathrm{Var}(\hat P_f)$ accounts for both the bias and variance of the estimator. ${\rm Cost}(\hat P_f)$ denotes the number of total model evaluations. In all numerical examples, we run both SuS and SDIS independently for 300 times to estimate the bias and variance of both estimators. 

In all examples, the parameters of the enhanced SDIS and the standard SDIS \cite{cheng2023rare} are set to $\sigma_1=3, n_s = 150, l=5$. In SuS, the sample size per level is set to $1000$, and the intermediate failure probability is set to $0.1$. The Matlab package of enhanced SDIS and standard SDIS is available at \url{https://github.com/KaiChengDM/Enhanced-SDIS/tree/main} and \url{https://github.com/KaiChengDM/SDIS}. The SuS package used in this work for comparison is available at \url{ https://github.com/ERA-Software/Overview/tree/main/Reliability%20Analysis%20Tools/3.%20Subset%20Simulation/SuS_Matlab}.

\subsection{ Two-dimensional example with multiple failure domains}
\label{example:2d}
In this example, we investigate the performance of  the enhanced SDIS with the two-dimensional LSF adapted from \cite{huang2022new}:
\begin{equation}
\begin{aligned}
 G(\bsx)= 5\left(4-2.1x_1^2 + \frac{ x_1^4}{3}\right)x_1^2 + 5x_1x_2 + 10( x_2^2-1)x_2^2 + 2.6,
 \label{eq:LSF2}
\end{aligned}
\end{equation}
where $x_1$ and $x_2$ are independent normal random variables with zero mean and standard deviation 0.05 and 0.18, respectively. We transform the LSF in Eq. \eqref{eq:LSF2} into standard normal space via isoprobabilistic transformation. The MCS reference value of failure probability with $10^{8}$ samples is $P_f = 3.71\times 10^{-5}$. 

\begin{figure}[htp]
  \centering
\includegraphics[width=1\textwidth,trim={20 295 0 295},clip]{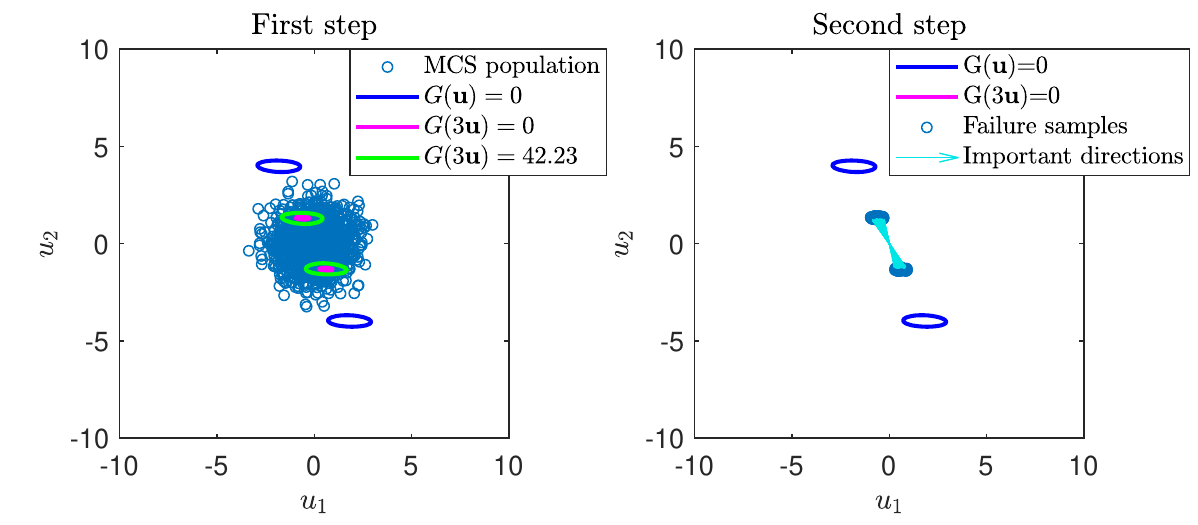}
  \caption{ Samples, important directions and intermediate LSFs of SDIS in standard normal sapce in problem \ref{example:2d}.}
  \label{fig:2d_example}
  \end{figure}

 Fig. \ref{fig:2d_example} depicts the limit state surface and a specific run of the enhanced SDIS for estimating the failure probability. It shows that the first auxiliary limit state surface $G(3\bsu)=0$ is much closer to the origin compared to the true one. However, we only obtain $17$ failure samples from the MCS population of size $10n_s$ drawn from standard normal PDF $\varphi_2(\bsu)$, and SuS is applied to estimate $ 
 P_{\sigma_1}$ in this case. In order to obtain $n_s$ failure samples in the intermediate failure domain $F_{\sigma_1,b_1}=\{\bsu\in \mathbb{R}^{n} :G(\sigma_1 \bsu)\leq b_1\}$, the failure threshold is set to $b_1=0.415$, and we obtain an estimate $\hat 
 P_{\sigma_1}= 0.012$ with only one intermediate step in SuS.
 Then, we estimate the ratio $\hat S_1=P_f/P_{\sigma_1}$, and a convergent estimator is obtained with only one additional step. The final estimate of failure probability is given by $\hat P_f= \hat P_{\sigma_1}\hat S_1$.
 
The reliability analysis results of the standard and enhanced SDIS and SuS are listed in Table \ref{tab:2d_example}, in which the mean of failure probability $\mathbb{E}(\hat P_f)$, the mean of the approximated CoV $\mathbb{E}(\hat{\delta}_{\hat P_f})$, the empirical CoV of the failure probability $\delta(\hat P_f)$, the mean of the total model evaluations $\mathbb{E}(N_t)$, and the relEff defined in Eq. \eqref{eq:releff} are provided. The approximated CoV of the enhanced SDIS is obtained by Eq. \eqref{eq:cov_sdis}, and that of SuS is obtained by assuming independence of the estimates of the conditional probabilities in different level \cite{au2001estimation,papaioannou2015mcmc}. The empirical CoV of the failure probability $\delta(\hat P_f)$ is obtained by repeating the entire analysis 300 times, hence this is the more realistic estimate of the CoV and is used to evaluate the relEff.  

Since the enhanced SDIS requires fewer intermediate steps, one can see that it provides much more robust results with fewer model evaluations than SuS, and the relative efficiency relEff of the enhanced SDIS is one magnitude larger than that of SuS. Additionally, it is found that the approximated CoVs of both methods underestimate the true ones due to the independence assumption. In this problem, standard SDIS is inefficient since the standard root finding algorithm implemented used within this method can only find one root in every important direction, and the estimator is severely biased. 

\begin{table}[hbt!]
\setlength{\belowcaptionskip}{1pt}
\caption{Reliability analysis results for example \ref{example:2d} (Reference value: $P_f = 3.71\times 10^{-5}$ )}
\centering
  \begin{tabular}{cccccc}
  \hline
    Method & $\mathbb{E}(\hat P_f)$ &  $\mathbb{E}(\delta_{\hat P_f})$ & $\delta(\hat P_f)$ & $\mathbb{E}(N_t)$ & relEff \\
  \hline
  SuS  &  $3.66\times 10^{-5}$ &  $ 0.30$  & $0.37$ & 5232  &  37.78 \\ Enhanced SDIS  &  $3.77\times 10^{-5}$ &   $0.15$ & 0.17 & 3491 & $255.49$\\
  Standard SDIS  &  $5.09\times 10^{-6}$ &   $0.11$ & 0.12 & 1249 & $2.89$ \\ 
 \hline
\end{tabular}
\label{tab:2d_example}
\end{table}

\subsection{Two-dimensional metaball function}
\label{example:metaball}
We consider the metaball LSF \cite{breitung2019geometry}:
\begin{equation*}
\begin{aligned}
 G(\bsu)= \frac{30}{\left(\frac{4(u_1+2)^2}{9}+\frac{u_2^2}{25}\right)^2+1} + \frac{20}{\left(\frac{(u_1-2.5)^2}{4}+\frac{(u_2-0.5)^2}{25}\right)^2+1}-5,
\end{aligned}
\end{equation*}
where $u_1$ and $u_2$ are standard normal random variables. The reference value of the failure probability obtained by MCS with $10^{8}$ samples is $P_f = 1.12\times 10^{-5}$. 

\begin{figure}[htp]
  \centering
\includegraphics[width=1 \textwidth,trim={40 300 1 300},clip]{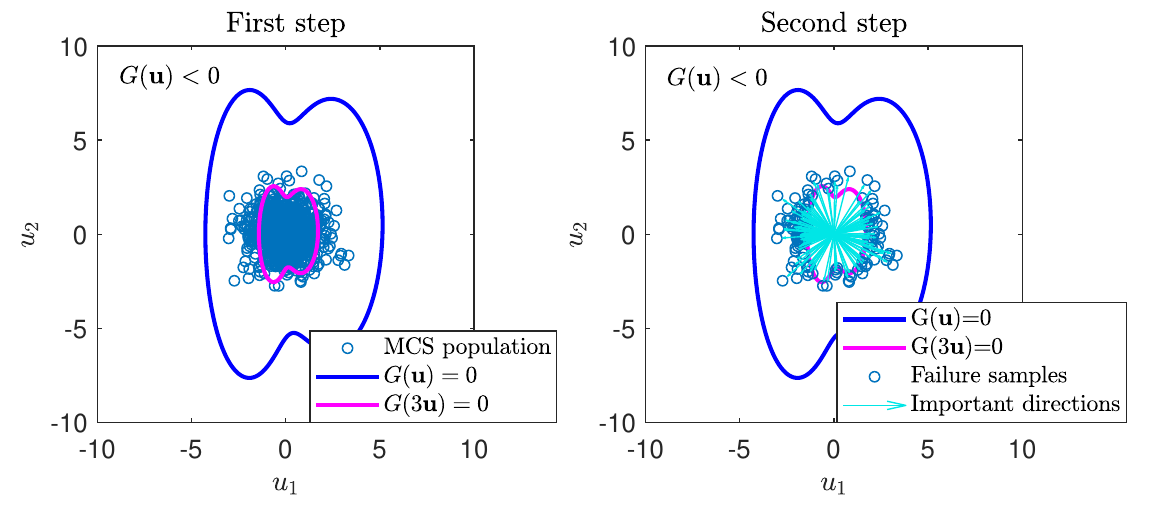}
  \caption{ Samples, important directions and intermediate LSFs of SDIS in problem \ref{example:metaball}.}
  \label{fig:metaball}
  \end{figure}

 Fig. \ref{fig:metaball} depicts the limit state surface and a specific run of the enhanced SDIS for estimating the failure probability in this example. Although there is only one failure domain, the failure boundary is irregular and highly nonlinear. In  the enhanced SDIS, the first auxiliary limit state surface $G(3\bsu)=0$ is much closer to the origin than the true one, and the first auxiliary failure probability is readily estimated with MCS. The $n_s=150$ failure samples in the MCS population define $n_s$ directional vectors $\{\bsa_i = \bsu_i/\Vert\bsu_i\Vert ,i=1,...,n_s\}$ following the first DIS density $h_{\sigma_1}(\bsa)$ in  the enhanced SDIS. By finding the roots along these directions with the Kriging model, we obtain an estimate $\hat S_1$ of $S_1$ in Eq. \eqref{eq:weight_mean}. Furthermore, $\sigma_2$ obtained by solving the optimization problem in Eq. \eqref{eq:opt_sigma} reduces to $1$ directly. Hence, one obtains a convergent estimation of failure probability with only one intermediate step in  the enhanced SDIS. By contrast, SuS gives poor result since the samples in the intermediate steps tend to be trapped in a local failure domain for this example \cite{breitung2019geometry,rashki2021sesc}.  
 
\begin{table}[hbt!]
\setlength{\belowcaptionskip}{1pt}
\caption{Reliability analysis results for example \ref{example:2d} (Reference value: $P_f = 1.12\times 10^{-5}$ )}
\centering
  \begin{tabular}{cccccc}
  \hline
    Method & $\mathbb{E}(\hat P_f)$ &  $\mathbb{E}(\delta_{\hat P_f})$ & $\delta(\hat P_f)$ & $\mathbb{E}(N_t)$ & relEff  \\
  \hline
  SuS  &  $1.14\times 10^{-5}$ &  $ 0.46$  & $2.93$ & 7751  &  $1.29$ \\ Enhanced SDIS  &  $1.07\times 10^{-5}$ &  $ 0.13$  & $0.15$ & 3562  &  $1163.58$  \\ Standard SDIS  &  $1.14\times 10^{-5}$ &  $ 0.14$  & $0.16$ & 3641  &  $982.85$  \\
 \hline
\end{tabular}
\label{tab:metaball}
\end{table}

The results are summarized in Table \ref{tab:metaball}. Both the standard and enhanced SDIS provide much more robust results with less model evaluations compared to SuS. Indeed, SuS degenerates to crude Monte Carlo simulation in this challenging problem. However, this problem has been specifically designed in \cite{breitung2019geometry} to cause problems for SuS, so the comparison may be unfair. Since there is only one root in every important direction, the standard SDIS is comparable to the enhanced SDIS in this example.

\subsection{Nonlinear oscillator}
\label{example:nonlinear_oscillator}
We test the performance of the enhanced SDIS method by assessing the reliability of a two-degree-of-freedom primary–secondary system under a white noise base
acceleration \cite{kiureghian1991efficient,bourinet2016rare}. The system failure occurs if the peak response of the secondary spring during the duration of the excitation
exceeds the threshold, and the corresponding LSF reads: 
\begin{equation}
\begin{aligned}
    g(\bsx)=F_s-3k_s\sqrt{\frac{\pi S_0}{4\zeta_s\omega_s^3}\left[\frac{\zeta_a\zeta_s}{\zeta_p\zeta_s\left(4\zeta_a^2+\theta^2\right)+\gamma\zeta_a^2}\frac{\left(\zeta_p\omega_p^3+\zeta_s\omega_s^3\right)\omega_p}{4\zeta_a\omega_a^4}\right]}
    \label{eq:oscillator}
\end{aligned}
\end{equation}
where $ \omega_{p}=\sqrt{k_{p}/m_{p}}, \omega_{s}=\sqrt{k_{s}/m_{s}}, \omega_{a}=(\omega_{p}+\omega_{s})/2, \zeta_{a}=\left(\zeta_{p}+\omega_{s}\right) \zeta_{s})/2, \gamma=m_{s}/m_{p}$ and $\theta=(\omega_{p}-\omega_{s})/\omega_{a}$. All the uncertain parameters are lognormally
distributed random variables and their distribution parameters are listed in Table \ref{tab:oscillator_parameter}.
 We transform the LSF in Eq. \eqref{eq:oscillator} into standard normal space via an isoprobabilistic transformation. A reference value of the ailure probability is obtained by MCS with $10^{8}$ samples as $P_f = 4.42\times 10^{-5}$. 

\begin{table}[hbt!]
\setlength{\belowcaptionskip}{1pt}
\caption{Random variables and their distribution parameters in Example \ref{example:nonlinear_oscillator} }
\centering
  \begin{tabular}{ccccccccc}
  \hline
 Variable & $m_p$  &  $m_s$ & $k_p$ & $k_s$ & $\zeta_p$ & $\zeta_s$ & $F_s$ & $S_0$\\
  \hline
  Mean  &  $1.5$ &   $0.01$ & $1$ & $0.01$ & $0.05$ & $0.02$ &  $22$ &  $100$
  \\  Cov & $0.1$  &  $0.1$ & $0.2$ & $0.2$ & $0.4$ & $0.5$ & $0.1$ & $0.1$ \\ 
 \hline
\end{tabular}
\label{tab:oscillator_parameter}
\end{table}

The results are listed in Table \ref{tab:oscillator_results}. They show that both the enhanced SDIS and SuS provide the desired results, but the enhanced SDIS outperforms the SuS in this example. For this challenging problem, the standard SDIS provides a severely biased estimate since there are multiple roots in several important directions, as shown in Figs. \ref{fig:Kriging_root2} and \ref{fig:Kriging_root3}. 

\begin{table}[hbt!]
\setlength{\belowcaptionskip}{1pt}
\caption{Reliability analysis results for example \ref{example:2d} (Reference value: $P_f = 4.42\times 10^{-5}$ )}
\centering
  \begin{tabular}{cccccc}
  \hline
    Method & $\mathbb{E}(\hat P_f)$ &  $\mathbb{E}(\delta_{\hat P_f})$ & $\delta(\hat P_f)$ & $\mathbb{E}(N_t)$ & relEff \\
  \hline
  SuS  &  $4.54\times 10^{-5}$ &  $ 0.31$  & $0.45$ & 5145 &  20.89 \\
  Enhanced SDIS  &  $4.16\times 10^{-5}$ &   $0.21$ & 0.40 & 5698 & $27.92$ \\
  Standard SDIS  &  $1.27\times 10^{-5}$ &   $0.19$ & 0.30 & 13412 & $3.28$ \\
 \hline
\end{tabular}
\label{tab:oscillator_results}
\end{table}

\subsection{High dimensional series system problem with two nonlinear branches}
\label{example:series_nonlinear}
In this section, we test the performance of the enhanced SDIS method by a nonlinear series system. The LSF is given by:
\begin{equation}
\begin{aligned}
  G(\boldsymbol{u})= {\rm min} \left\{\beta-\frac{1}{\sqrt{n}}\sum_{i=1}^n u_i + (u_1-u_2)^2/10 , \beta+\frac{1}{\sqrt{n}}\sum_{i=1}^n u_i + (u_1-u_2)^2/10  \right\}
\end{aligned}
\end{equation}

\begin{figure}[htp]
  \centering
\includegraphics[width=1\textwidth,trim={10 270 10 280},clip]{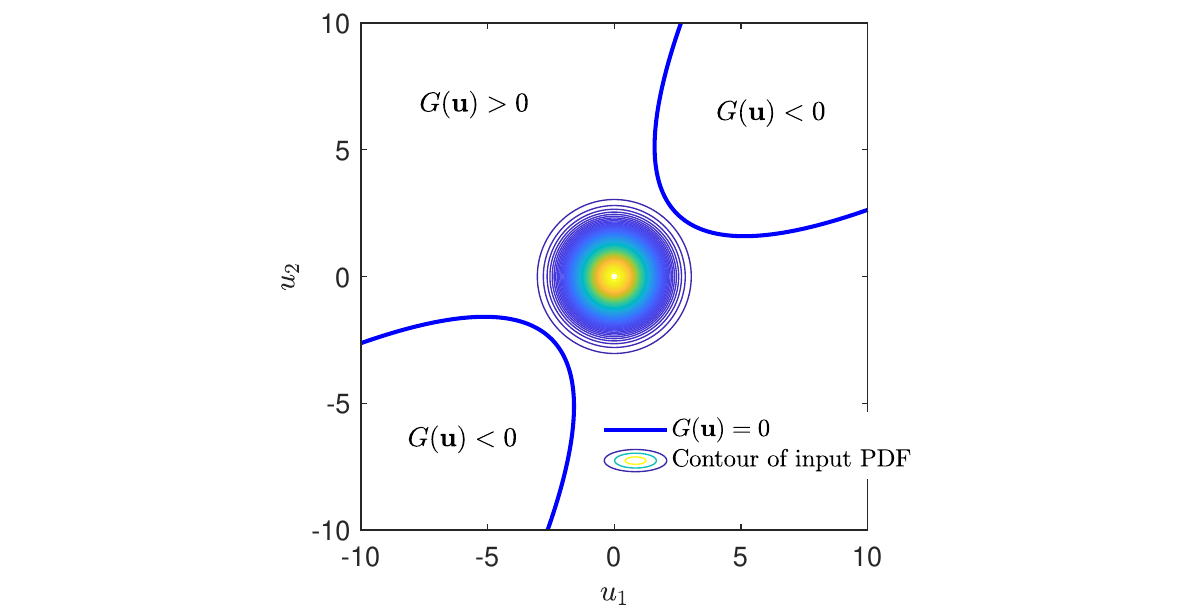}
  \caption{ Two dimensional visualized LSF of example \ref{example:series_nonlinear}.}
  \label{fig:nonlinear_series}
  \end{figure}

In this example, we set $\beta=3.5$, and the reference failure probability is $2.92\times 10^{-4}$, which is independent of the number of input variables $n$. To better understand the shape the LSF considered here, Fig. \ref{fig:nonlinear_series} visualizes the limit state surface for $n=2$. 
One can see that there are two failure domains in opposite sides of the origin. 

 We test the performance of the enhanced SDIS for varying input dimensions $n$. The results for all methods are reported in Table \ref{tab:nonlinear_series}. All methods give reasonable accuracy and efficiency in all settings. The enhanced SDIS outperforms SuS for $n=10$. However, the performance of the enhanced SDIS degenerates with the rise of input dimension, and SuS is superior for $n=100$ and $n=1000$. Overall, the enhanced SDIS is comparable to SuS in this problem and it always outperforms the standard SDIS. 

 \begin{table}[htbp]
 \caption{Reliability analysis results for example \ref{example:series_nonlinear} }
  \centering
  \begin{tabular}{cccccc}
  \toprule $n=10$
  &  $\mathbb{E}(\hat P_f)$ & $\mathbb{E}(\hat \delta_{\hat P_f})$  & $\delta_{\hat P_f}$ & $\mathbb{E}(N_t)$ & relEff\\
 \midrule
SuS &  $2.95\times 10^{-4}$ & $0.25$  &  0.28 &  4404 & 9.68\\ Enhanced SDIS & $2.93\times 10^{-4}$  & 0.17 &  0.26  & 2922 & 17.39 \\ Standard SDIS & $2.94\times 10^{-4}$  & 0.19 &  0.25  & 3936 & 13.81\\
  \bottomrule
  \end{tabular}
  \begin{tabular}{cccccc}
  \toprule $n=100$
  &  $\mathbb{E}(\hat P_f)$ & $\mathbb{E}(\hat \delta_{\hat P_f})$  & $\delta_{\hat P_f}$ & $\mathbb{E}(N_t)$ & relEff\\
 \midrule
SuS &  $2.94\times 10^{-4}$ & $0.25$  &  0.29 &  4405 & 9.24\\ Enhanced SDIS & $2.84\times 10^{-4}$  & 0.24 &  0.31  & 4531 & 8.15 \\  Standard SDIS & $2.88\times 10^{-4}$  & 0.25 &  0.29  & 6829 & 6.11 \\
  \bottomrule
  \end{tabular}
  \begin{tabular}{cccccc}
  \toprule $n=1000$
  &  $\mathbb{E}(\hat P_f)$ & $\mathbb{E}(\hat \delta_{\hat P_f})$  & $\delta_{\hat P_f}$ & $\mathbb{E}(N_t)$ & relEff\\
 \midrule
SuS &  $3.03\times 10^{-4}$ & $0.25$  &  0.29 &  4397 & 8.69 \\Enhanced SDIS & $2.87\times 10^{-4}$  & 0.27 &  0.31  & 5260 & 7.06 \\ Standard SDIS & $2.73\times 10^{-4}$  & 0.28 &  0.29  & 8946 & 5.01 \\
  \bottomrule
  \end{tabular}
  \label{tab:nonlinear_series}
\end{table}

\subsection{High dimensional nonlinear example}
\label{example:high_nonlinear}
We now test the performance of the enhanced SDIS method with the following LSF in standard normal input space \cite{fujita1988}:
\begin{equation}
 G(\bsu)= C_a+\sum_{i=1}^n {\rm ln}(\Phi^{-1}(-u_i)).
 \label{eq:highly_nonlinear}
\end{equation}

\begin{figure}[htp]
  \centering
\includegraphics[width=1\textwidth,trim={10 280 1 290},clip]{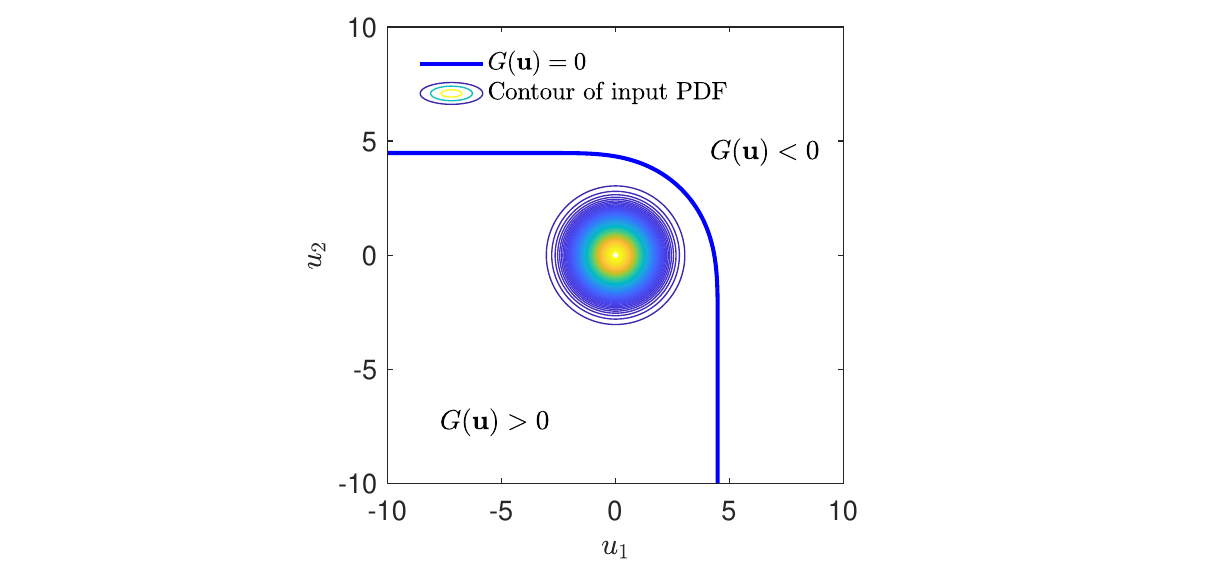}
  \caption{ Two dimensional visualized LSF of example \ref{example:high_nonlinear}.}
  \label{fig:nonlinear}
  \end{figure}

The function in Eq. \eqref{eq:highly_nonlinear} is highly nonlinear. The reference value of probability of failure is $P_f = 1-F_Y(C_a)$, where $Y$ is a Gamma distributed random variable with shape parameter $n$ and scale parameter  $\lambda=1$. 
In this example, we vary the input dimension from $n=10$ to $1000$. We adjust the value of $C_a$ to fix $P_f=5\times 10^{-5}$ for different $n$. The visualized limit state surface for $n=2$ is depicted in Fig. \ref{fig:nonlinear}. 
One can see that the 2-dimensional limit state surface is nonlinear and it has a convex safe domain. 

 The results are summarized in Table \ref{tab:nonlinear}. Enhanced SDIS yields good results for various input dimensions $n$.
 It outperforms SuS for $n=10$. With increasing input dimension, the performance of the enhanced SDIS degenerates gradually, and it is inferior to SuS for $n=100$ and $n=1000$. In addition, it is observed that the enhanced SDIS always outperforms the standard SDIS considerably, which is due to the increased efficiency of the proposed Kriging-based root-finding algorithm compared to the one applied in standard SDIS.
 
 \begin{table}[htbp]
 \caption{Reliability analysis results for example \ref{example:high_nonlinear} }
  \centering
  \begin{tabular}{cccccc}
  \toprule $n=10$
  &  $\mathbb{E}(\hat P_f)$ & $\mathbb{E}(\hat \delta_{\hat P_f})$  & $\delta_{\hat P_f}$ & $\mathbb{E}(N_t)$ & relEff\\
 \midrule
SuS &  $5.04\times 10^{-5}$ & $0.27$  &  0.32 &  5094 & 38.40\\ Enhanced SDIS & $4.81\times 10^{-5}$  & 0.16 &  0.24  & 2070 & 170.76 \\ Standard SDIS & $4.99\times 10^{-5}$  & 0.17 &  0.25  & 3254 & 102.25 \\ 
  \bottomrule
  \end{tabular}
  \begin{tabular}{cccccc}
  \toprule $n=100$
  &  $\mathbb{E}(\hat P_f)$ & $\mathbb{E}(\hat \delta_{\hat P_f})$  & $\delta_{\hat P_f}$ & $\mathbb{E}(N_t)$ & relEff\\
 \midrule
SuS &  $4.94\times 10^{-5}$ & $0.27$  &  0.326 &  5105 & 37.54\\ Enhanced SDIS & $4.76\times 10^{-5}$  & 0.22 &  0.44  & 4211 & 26.53\\ Standard SDIS & $4.88\times 10^{-5}$  & 0.22 &  0.45  & 6905 & 15.18\\ 
  \bottomrule
  \end{tabular}
  \begin{tabular}{cccccc}
  \toprule $n=1000$
  &  $\mathbb{E}(\hat P_f)$ & $\mathbb{E}(\hat \delta_{\hat P_f})$  & $\delta_{\hat P_f}$ & $\mathbb{E}(N_t)$ & relEff\\
 \midrule
SuS &  $5.13\times 10^{-5}$ & $0.28$  &  0.339 &  5089 & 32.39\\ Enhanced SDIS & $4.91 \times 10^{-5}$  & 0.25 &  0.51  & 4988 & 16.09\\ Standard SDIS & $4.70 \times 10^{-5}$  & 0.25 &  0.50  & 11851 & 7.48\\
  \bottomrule
  \end{tabular}
  \label{tab:nonlinear}
\end{table}

\section{Conclusion}
\label{sec:conclusion}

We presented an enhanced sequential directional importance sampling (SDIS) method for structural reliability analysis. The main contributions that lead to improved performance are twofold. First, we proposed to estimate the first integral in SDIS with Subset Simulation (SuS) in problems with narrow failure domains. Second, we developed a Kriging-based active learning algorithm that is able to identify multiple roots within a specified search interval and has increased efficiency compared to the root-finding algorithm used in the standard SDIS.
  
  Several benchmarks were investigated to compare the performance of the enhanced SDIS to that of standard SDIS and SuS. The results demonstrated that the enhanced SDIS outperforms the standard one in terms of efficiency and robustness. It is a versatile reliability analysis method, which can potentially address various challenging reliability analysis problems, i.e., problems with highly nonlinear limit state function, low magnitude of failure probability, multiple disjoint failure domains and high-dimensional input parameter. As with standard SDIS, the performance of the enhanced SDIS still degenerates with increasing input variable dimension. This is attributed to the fact that the variance of the conditional failure probability ${\rm Pr}\left(G(\sigma R\bsa)\leq 0\right)$ rises gradually when the input dimension increases, thus requiring more intermediate steps to achieve convergence in SDIS. In problems with more than 100 input variables, SuS still outperforms the enhanced SDIS. 

 \appendix
\section{The ordinary Kriging model}
\label{sec:Kriging}
 In the ordinary Kriging model, the one-dimensional computational model  $y=G(\sigma r\bsa)$ in direction $\bsa$ is assumed to be a Gaussian process  
   \begin{equation*}
   Y(r)= \beta_0 + Z(r),
   \label{assumption}
   \end{equation*}
   where $\beta_0\in\R^{m} $ is the mean of $Y(r)$, and  $Z(r)$ is a stationary Gaussian process with zero mean and covariance function given by
   \begin{equation*}
 \text{Cov}\left(Z(r),Z(\hat{r})\right)=\sigma_0^2k(r,\hat{r};{\theta}),
   \end{equation*}
   
  \noindent where  $\sigma_0^2$ is the variance of $Z(r)$, and $k(r,\hat{r};\theta)$ represents the spatial correlation between $r$ and $\hat{r}$ with the correlation length controlled by the hyper-parameter $\theta$. 
 
  Under the Guassian process assumption, the joint distribution of $Y(r)$ and $\boldsymbol{Y}=[Y(r^{(1)}),\ldots,Y(r^{(M)})]^\mathrm{T}$ at the sampled locations is multi\-variate Gaussian, namely
     \begin{equation*}
     \left[\begin{array}{cc}Y(r) \\ \boldsymbol{Y} \end{array}\right] \sim \mathcal {N}\left(\left[\begin{array}{cc} \beta_0 \\ 
\beta_0\boldsymbol{F}\end{array}\right],\sigma_0^2\left[\begin{array}{cc} 1 & \boldsymbol{k}^\mathrm{T} (r)\\ \boldsymbol{k}(r) & \boldsymbol{K} \end{array} \right] \right),
    \end{equation*}
     where $\boldsymbol{F}:= [1,...,1]^\mathrm{T}\in \mathbbm{R}^{M}$ and 
    \begin{eqnarray*}
    \begin{aligned}
     &\boldsymbol{K}
     := k(r^{(i)},{r^{(j)}};\theta),
      i,j=1,\ldots,M, \\
      &\boldsymbol{k}(r):= k(r^{(i)},{r};{\theta}),  i=1,\ldots,M.
      \end{aligned}
    \end{eqnarray*}
     
     Given the observation values $\boldsymbol{Y}=[G(\sigma_i r^{(1)}\bsa ),...,G(\sigma_i r^{(M)}\bsa )]^{\rm T}$, 
     the predictive distribution of $Y(r)$ is given by
     \begin{equation*}
      \hat Y(r) \sim \mathcal {N}\left(\mu(r), s^2(r)\right),
     \label{predictor}
     \end{equation*}
    where the mean and variance are given by 
    \begin{eqnarray*}
    \begin{aligned}
    \mu(r) &= \beta_0+\boldsymbol{k}^\mathrm{T}(r)\boldsymbol{K}^{-1}(\boldsymbol{Y}- \beta_0\boldsymbol{F}), \\
     s^2(r) &= \sigma_0^2 \left(1- \boldsymbol{k}^\mathrm{T}(r) \boldsymbol{K}^{-1}\boldsymbol{k}(r)+ \boldsymbol{u}^{\rm T}(r)(\boldsymbol{F}^\mathrm{T}\boldsymbol{R}^{-1}\boldsymbol{F})^{-1}\boldsymbol{u}(r)\right).
   \end{aligned}
    \end{eqnarray*}
in which $\boldsymbol{u}(r)=\boldsymbol{F}^\mathrm{T}\boldsymbol{K}^{-1}K(r)-1$. 

In ordinary Kriging model, the mean $\beta_0$ and variance $\sigma_0^2$ are given by 
 \begin{eqnarray*}
    \beta_0 &=& (\boldsymbol{F}^\mathrm{T}\boldsymbol{K}^{-1}\boldsymbol{F})^{-1}\boldsymbol{F}^\mathrm{T}\boldsymbol{K}^{-1}\boldsymbol{Y},
     \label{beta} \\
     \sigma_0^2 &=& \frac1{M}(\boldsymbol{Y}- \beta_0\boldsymbol{F})^{\rm T}\boldsymbol{K}^{-1}(\boldsymbol{Y}- \beta_0\boldsymbol{F}).
    \label{sigma2}
    \end{eqnarray*}

The hyper-parameter $\theta$ is determined with maximum likelihood estimation method, which is equivalent to minimizing the following objective function:
\begin{equation*}
     \ell(\theta)=M\ln \sigma_0^2 + \ln\left[{\rm det\boldsymbol{K}}\right].
   \label{reducedlikelihood}
   \end{equation*}

\section{Acknowledgements}
 This work was supported by the Alexander von Humboldt Foundation.

\bibliographystyle{plain}
\bibliography{Ref}
\end{document}